\documentclass{webofc}
\usepackage[varg]{txfonts}   % Web of Conferences font
\usepackage{color}
\usepackage[normalem]{ulem}
\usepackage{floatrow}
\begin{document}
\title{Beyond-mean-field calculations of allowed and first-forbidden $\beta^-$ decays of $r$-process waiting-point nuclei}

\author{\firstname{Caroline} \lastname{Robin}\inst{1,2}\fnsep\thanks{\email{c.robin@gsi.de}} \and
        \firstname{Elena} \lastname{Litvinova}\inst{3,4}\fnsep \and
        \firstname{Gabriel} \lastname{Mart\'inez-Pinedo}\inst{1,5,6}\fnsep        
}

\institute{GSI Helmholtzzentrum f\"ur Schwerionenforschung, Planckstra\ss e 1, 64291 Darmstadt, Germany
\and
           Fakult\"at f\"ur Physik, Universit\"at Bielefeld, D-33615, Bielefeld, Germany
\and
           Department of Physics, Western Michigan University, Kalamazoo, Michigan 49008, USA
\and
           National Superconducting Cyclotron Laboratory, Michigan State University, East Lansing, Michigan 48824, USA
\and
           Institut f\"ur Kernphysik (Theoriezentrum), Fachbereich Physik, Technische Universit\"at Darmstadt, Schlossgartenstra\ss e 2, 64298 Darmstadt, Germany
\and           
          Helmholtz Forschungsakademie Hessen f\"ur FAIR, GSI Helmholtzzentrum f\"ur Schwerionenforschung, Planckstra\ss e 1, 64291 Darmstadt, Germany
          }

\abstract{%
$\beta$-decay rates of neutron-rich nuclei, in particular those located at neutron shell closures, play a central role in simulations of the heavy-element nucleosynthesis and resulting abundance distributions. 
We present $\beta$-decay half-lives of even-even $N=82$ and $N=126$ $r$-process waiting-point nuclei calculated in the approach based on relativistic quasiparticle random phase approximation with quasiparticle-vibration coupling.
The calculations include both allowed and first-forbidden transitions. In the $N=82$ chain, the quasiparticle-vibration coupling has an important impact close to stability, as it increases the contribution of Gamow-Teller modes and improves the agreement with the available data. In the $N=126$ chain, we find the decay to proceed dominantly via first-forbidden transitions, even when the coupling to vibrations is included.
}
\maketitle
%%%%%%%%%%%
\section{Introduction}
Modeling the $r$-process nucleosynthesis requires a huge amount of nuclear physics input, and in particular, the knowledge of $\beta$-decay rates for thousands of neutron-rich nuclei. Due to their very short lifetimes, most of these nuclei are not accessible experimentally and astrophysical simulations thus heavily rely on theoretical methods, which should be as universal and reliable as possible.
\\
As of today the only approach able to provide global sets of $\beta$-decay rates is known as Quasiparticle Random Phase Approximation (QRPA). Such method, which corresponds to the small amplitude limit of the time-dependent mean-field approach, has the advantage to be applicable to a wide range of masses at relatively low numerical cost. QRPA, however, suffers well-known limitations related to a simplified treatment of nuclear correlations which restrict its reliability and predictive power, in particular in unexplored regions of the nuclear chart.
Currently, three global sets of $\beta$-decay rates are available, which have been calculated within different QRPA frameworks. The earlier set based on the finite range droplet model and schematic interaction \cite{Moeller2003}, and two more recent self-consistent and fully microscopic QRPA calculations based on a relativistic Langrangian \cite{Marketin2016} and non-relativistic Skyrme functional \cite{Ney2020}.
Overall the predictions of these three QRPA frameworks disagree to some degree, in particular on the contribution of the FF transitions in different mass regions, and on the order of magnitude of the total half-lives above $N>126$ which are predicted to be substantially lower by Ref. \cite{Marketin2016} than by the other two approaches. This large spread in the prediction of $\beta$-decay half-lives is an important issue as it can yield large uncertainties in the predictions of elemental abundances \cite{Shafer2016}. For example, it was found in Refs.~\cite{Marketin2016,Eichler_2015} that the shortenings of half-lives in the heavy region yields a broadening of the third $r$-process abundance peak towards lower masses.
\\
While further studies are needed to fully understand the origin of the discrepancies, it is desirable to improve the reliability of $\beta$-decay rates by developing theoretical methods which incorporate a more precise description of nucleonic correlations, while remaining applicable to a wide range of masses. In this paper we use the approach based on charge-exchange relativistic QRPA (RQRPA) that is extended to account for the coupling between nucleons and collective nuclear vibrations, as first developed in Ref. \cite{Robin2016}, and apply it to the description of allowed and first-forbidden $\beta$ decays of $N=82$ and $N=126$ $r$-process waiting-point nuclei. 
%%%%%%%%%%%%
\section{Quasiparticle-vibration coupling effect on allowed and first-forbidden $\beta^-$ decay of $N=82$ and $N=126$ nuclei.}
\begin{figure}[h]
\floatbox[{\capbeside\thisfloatsetup{capbesideposition={left,top},capbesidewidth=3.5cm}}]{figure}[\FBwidth]
{\caption{$\beta$-decay half-lives of $N=82$ (left) and $N=126$ (right) nuclei. We show the results obtained within RQRPA (green triangles) and RQRPA+QVC (red diamond) in the allowed GT approximation (empty symbols) and including FF transitions (full symbols). When no symbol is shown, the half-life is predicted infinite. The experimental data is taken from \cite{data}. }\label{fig-1}}
{\includegraphics[width=9.cm]{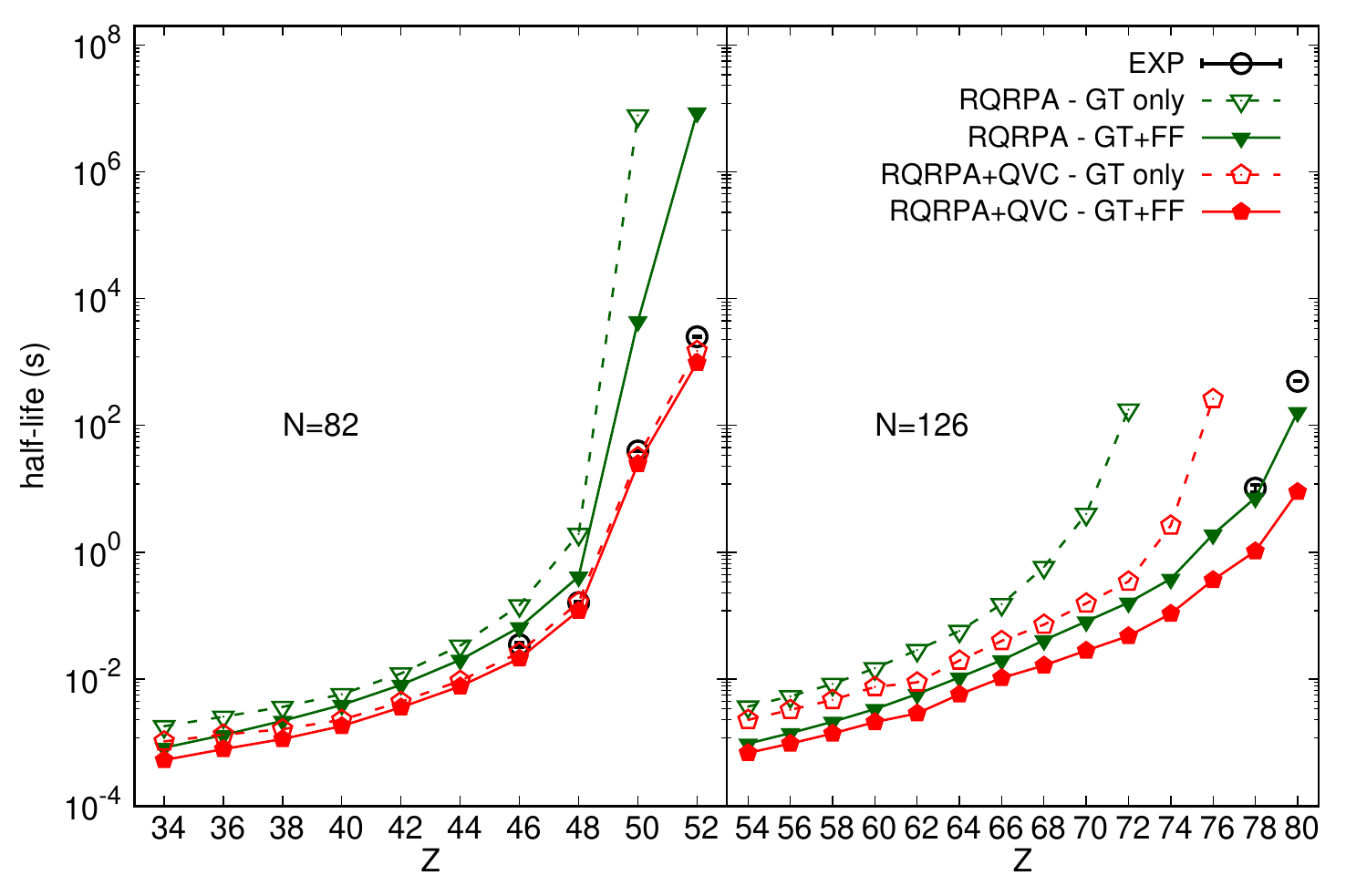}}
\end{figure}

In this work we calculate the $\beta$-decay rates $\lambda$ including both allowed Gamow-Teller (GT) and first-forbidden (FF) transitions according to Ref.~\cite{BEHRENS1971,Marketin2016}:
\begin{equation}
\lambda =  \frac{\mbox{ln 2} }{K}   \sum_{n , \Omega_n < \Delta M_{nH}} f(\Omega_n, Z+1) \; .
\label{eq:rate}
\end{equation}
In Eq. \ref{eq:rate}, $K=1644 \pm 2 $ s \cite{K}, the index $n$ denotes the states of the daughter nucleus $(Z+1,N-1)$ accessible via $\beta$-decay, $\Omega_n$ is the excitation energy of these states with respect to the parent ground state, $\Delta M_{nH}\simeq 0.78$ MeV is the neutron-Hydrogen mass difference and $f(\Omega_n, Z+1)$ is the so-called integrated shape function which takes the form
\begin{eqnarray}
f(\Omega_n, Z+1) = \int_1^{W_n} dW W \sqrt{W^2-1} (W_n -W)^2 F(Z+1, W) \left[ C_{allowed}(W) +  C_{FF}(W) \right]\; ,
\end{eqnarray}
where $W = E_e/ m_e c^2$ denotes the electron energy in unit of the electron mass, $W_n \equiv (M_i - M_n)/(m_e c^2) = [(m_n-m_p) -\Omega_n]/(m_e c^2)$ is the mass difference between initial and final states, $F(Z+1, W) $ is the Fermi function \cite{BEHRENS1971}, and $C_{X}(W)$ are the allowed and FF shape factors. In the case of allowed transitions $C_{allowed}(W)$ coincides with the GT reduced transition probabilities,
while $C_{FF}(W)$ takes a complicated expression which depends on energy and transition amplitudes of $0^-, 1^-$ and $2^-$ FF modes (for more details see \cite{Marketin2016,Robin2021}).
In this work we adopt the bare value of the weak axial coupling constant $g_A\simeq -1.27641$ \cite{ga} which appears in the definition of $C_X(W)$. 
\\
\\
The nuclear transition amplitudes are obtained within the approach based on the relativistic QRPA which is extended to account for the coupling between single quasiparticles and collective nuclear vibrations. For details on the formalism and numerical scheme, see Refs. \cite{Robin2016,Robin2021}. In this work we use the NL3 parametrization of the meson-nucleon Lagrangian \cite{NL3} for the particle-hole channel of the static interaction and a monopole-monopole force for the isovector pairing channel \cite{Litvinova2008}. The isoscalar pairing is only included dynamically via quasiparticle-vibration coupling and is not present in the static interaction \cite{Robin2021}.
\\
\\
We show in Fig.~\ref{fig-1} the $\beta$-decay half-lives $T_{1/2} =\mbox{ln 2} /\lambda$ of $N=82$ and $N=126$ neutron-rich nuclei obtained without and with quasiparticle-vibration coupling (QVC). The empty symbols show the half-lives obtained in the GT allowed approximation while the full symbols are the half-lives obtained with the contribution of FF transitions.
%
%\begin{figure}[h]
%\centering
%\includegraphics[width=9.2cm]{hlives_N82_N126.pdf}
%\caption{$\beta$-decay half-lives of $N=82$ (left) and $N=126$ (right) nuclei. We show the results obtained within RQRPA (green triangles) and RQRPA+QVC (red diamond) in the allowed GT approximation (empty symbols) and including the contribution of FF transitions (full symbols). When no symbol is shown, the half-life is predicted to be infinite. The experimental data is taken from \cite{data}. }
%\label{fig-1}       
%\end{figure}
%
%
\\
In both isotonic chains, the QVC correlations, which fragment the transition strength distributions \cite{Robin2016}, lead to a decrease of the half-lives which can be of several orders of magnitude compared to RQRPA. The effect is typically stronger close to stability, in particular in the $N=82$ chain, as the decay $Q$ value becomes small and is populated by very few states. In that case, the presence of correlations is crucial and leads to a much better agreement with the experimental data, although the decrease can be slightly too strong in some cases. Far from stability, however, the $Q$ value is large so that the details of the transition strength distributions matter less and the effect of correlations on the half-lives becomes small.
In the $N=126$ chain the decrease of the half-lives due to QVC is of one order of magnitude at most when the FF transitions are included. In fact, in this case, RQRPA is closer to the data for $Z=78$ and $Z=80$. The reader should remember, however, that we did not include any quenching of the GT or FF transition matrix elements. 
\begin{figure}[h!]
\centering
\includegraphics[width=\textwidth]{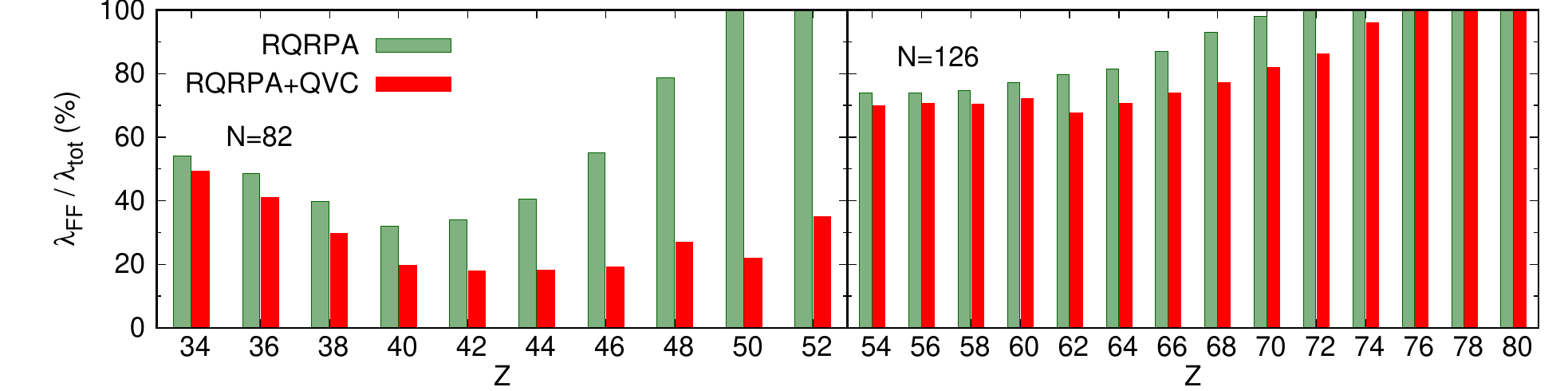}
\caption{Contributions of FF transitions to the total $\beta$-decay rate of $N=82$ (left) and $N=126$ (right) nuclei, obtained within RQRPA (green boxes) and RQRPA+QVC (red boxes).}
\label{fig-2}       
\end{figure}
\\
In Fig.~\ref{fig-2} we show the contribution of FF transitions to the total decay rate (in percent). 
In the $N=82$ isotonic chain, the contribution of FF modes increases far from stability from $Z=40$ to $Z=34$ where it reaches about $50-55 \%$ within both RQRPA and RQTBA. Examining the results in more details, we find that these FF transitions mainly occur via $0^-$ and $1^-$ modes. 
Going towards stability, the weight of FF transitions within RQRPA increases again until it reaches $\simeq 100 \%$ in $Z=50-52$. In these nuclei the proton $1g_{9/2}$ is fully occupied and thus, the main low-energy GT transition $1g_{7/2} \rightarrow 1g_{9/2}$ is blocked. The decay then proceeds solely via $2^-$ FF transitions (mostly $1h_{11/2} \rightarrow 1g_{7/2}$).
With QVC the contribution of FF modes remains below $\simeq 35\%$, which is of the same order of what was found in Shell-Model calculations \cite{SM2}.
Although the proton $1g_{9/2}$ is also filled for $Z\geq 50$, GT modes caused by transitions to other levels
are lowered in energy due to the correlations, and contribute significantly to the decay due to the phase space.
In the $N=126$ chain, FF transitions dominate compared to GT modes. In RQRPA we find a contribution of $\simeq 75 \%$ for $Z=54-58$ which then increases to $100 \%$ close to stability as the proton $1h_{11/2}$ sub-shell is filled. The trend remains similar when QVC correlations are included, apart from some variations. 
In this chain, Shell-Model studies have found a steeper increase of the FF contributions, from $\simeq 20-30 \%$ to $\simeq 60-75 \%$ in the range $Z=66-72$ \cite{SM1,SM2}.
Looking closer at the different forbidden transitions, we find that $0^-$ operators are responsible for the FF decay by $> 75 \%$ along the chain, within RQRPA+QVC. Among those, the relativistic component ($\gamma_5$ operator) appear to be more important than the spin-dipole component. 
\section{Conclusion}
We have calculated $\beta^-$-decay rates of $N=82$ and $N=126$ isotonic chains within an extension of RQRPA accounting for QVC, including GT and FF transitions. We find the effect of QVC to be important near stability, in particular in the $N=82$ chain, where it increases the contribution of GT modes. In $N=126$ isotones, the FF decay remains dominant, even when QVC is included. 
In a future work we will examine the role of FF transitions in other magic isotonic and isotopic chains, and compare to different theoretical frameworks \cite{Robin2021}.
\\ \\
This work is supported by ERC-885281-KILONOVA Advanced Grant and US-NSF Career Grant PHY-1654379, and used resources of the National Energy Research
Scientific Computing Center, a DOE Office of Science User Facility supported by the Office of Science of the U.S. Department of Energy under Contract No. DE-AC02-05CH11231.

%
% BibTeX or Biber users please use (the style is already called in the class, ensure that the "woc.bst" style is in your local directory)
 \bibliography{NIC}
%\begin{thebibliography}{}
%%
%% and use \bibitem to create references.
%Journal Author, Journal \textbf{Volume}, page numbers (year)
%Book Author, \textit{Book title} (Publisher, place, year) page numbers
%\end{thebibliography}

\end{document}